# Crossing the Architectural Barrier: Evaluating Representative Regions of Parallel HPC Applications

Alexandra Ferrerón
Universidad de Zaragoza, Spain
ferreron@unizar.es

Radhika Jagtap
ARM Ltd., U.K.
radhika.jagtap@arm.com

Sascha Bischoff
ARM Ltd., U.K.
sascha.bischoff@arm.com

Roxana Rușitoru
ARM Ltd., U.K.
roxana.rusitoru@arm.com

*Abstract*—Exascale computing will get mankind closer to solving important social, scientific and engineering problems. Due to high prototyping costs, High Performance Computing (HPC) system architects make use of simulation models for design space exploration and hardware-software co-design. However, as HPC systems reach exascale proportions, the cost of simulation increases, since simulators themselves are largely single-threaded. Tools for selecting representative parts of parallel applications to reduce running costs are widespread, e.g., BarrierPoint achieves this by analysing, in simulation, abstract characteristics such as basic blocks and reuse distances. However, architectures new to HPC have a limited set of tools available.

In this work, we provide an independent cross-architectural evaluation on real hardware—across Intel and ARM—of the BarrierPoint methodology, when applied to parallel HPC proxy applications. We present both cases: when the methodology can be applied and when it cannot. In the former case, results show that we can predict the performance of full application execution by running shorter representative sections. In the latter case, we dive into the underlying issues and suggest improvements. We demonstrate a total simulation time reduction of up to 178x, whilst keeping the error below 2.3% for both cycles and instructions.

## I. Introduction

High-Performance Computing (HPC) has changed the way we conduct scientific research by enabling complex problem-space exploration. Science and engineering have evolved quickly thanks to the computational power of supercomputers, which are able to compute billions of operations per second. The next HPC challenge is achieving an *exaflop*: $10^{18}$ FLoating-point Operations Per Second (FLOPS) in a 20 MW power budget [1]. However, current technology is not at the level necessary to accommodate the power and computational requirements for *exascale computing*. Meeting this target requires innovation in technology, architecture, software and programmability. Power dissipation, interconnects and memory bandwidth, among others, are fundamental areas that need to be tackled to reach the exascale landmark.

Due to practical and economic reasons, it is impossible to prototype every design alternative in order to evaluate novel hardware or to characterise the behaviour of new applications. Thus, the design of next-generation HPC systems needs to happen in tandem with fast, detailed and accurate application simulation. One of the main issues faced by HPC system designers is the simulation infrastructure, as accelerators and heterogeneity have added even more complexity to the already intricate models, resulting in exceedingly long simulation times. Simulation frameworks, such as *gem5* [2], are able to model different micro-architectures in great detail, but running full HPC applications is infeasible, due to the large simulation times involved; simulating the detailed behaviour of a workload that runs for as little as one second can take several hours [3].

In light of the aforementioned simulation issues, analysing applications to obtain representative sections that model the entire application behaviour is critical in order to confine simulation to the essential parts, only. This allows designers and researchers to reduce the total execution time without compromising the representativeness of results. In this context, many methodologies have been proposed [4], [5], [6], [7]. Sampling methodologies, such as SimPoint [4] and BarrierPoint [7], are very popular because of their effectiveness in reducing simulation time without sacrificing accuracy. They rely on a preliminary application analysis, observing high-level abstract characteristics, such as control flow (i.e., basic blocks) and memory patterns (i.e., reuse distances). This analysis is only available on a limited number of architectures.

In this paper, we evaluate whether the above-mentioned abstract characteristics are suitable for extracting the representative sections of HPC applications across different architectural features, such as vector capabilities and Instruction Set Architectures (ISAs), whilst running on real hardware. We use the BarrierPoint methodology, and extract representative phases of HPC applications on x86_64. Then, we investigate if the selected phases are able to capture the intrinsic behaviour of the programs for both x86_64 and ARMv8 architectures. This process allows for workloads to be evaluated on novel, upcoming HPC platforms. A main differentiation between prior work and our evaluation is that the original BarrierPoint methodology was evaluated in simulation, while we use real hardware. In addition, we offer an initial exploration into the cross-architectural applicability of BarrierPoint.

This work makes the following contributions:

1) We provide an independent evaluation of the BarrierPoint methodology when applied to a diverse set of state-of-the-art HPC proxy applications across two different architectures: x86_64 and ARMv8, on real hardware.
2) We show that a high level of abstraction can be used to identify the representative parts of HPC applications across both ISAs.
3) We look into architectural features (vector capabilities), and evaluate their impact on the representativeness of the selected phases, obtaining errors similar to the non-vectorised cases.

Our results show that the performance of the entire application (in terms of cycles and instructions) can be estimated with an error below 2.3%, for both architectures (x86_64 and ARMv8). These estimations are based on the representative regions identified on x86_64 and show the cross-architectural applicability of the methodology. Similar results are obtained when considering vectorised versions of the code: selected regions of vectorised binaries on x86_64 are representative across x86_64 and ARMv8. We achieve this high accuracy running as little as 0.6% of the total workload, and a maximum of 39%, which significantly reduces simulation time and enables us to increase the HPC proxy application input size (data set or number of threads). However, we find that the methodology presents some limitations, and it cannot be applied to some of the applications explored.

The rest of this paper is organised as follows. Section II reviews the related work and it contrasts it to our own. Section III provides a brief background about vector capabilities on Intel and ARM. Section IV describes the evaluation setup. Section V details the cross-architectural evaluation methodology. Section VI presents our evaluation results. Section VII concludes and summarises our key contributions. Finally, future work is presented in Section VIII.

Throughout this paper, we refer to the respective Intel and ARM architectures by their name, as follows: x86_64 refers to the 64-bit Intel architecture, whilst ARMv8 refers to the 64-bit ARM architecture.

## II. RELATED WORK

To the best of our knowledge, we are the first to evaluate if a sampling technique reliably identifies representative sections of HPC applications across different architectural features. Our contributions enable researchers to explore ARM-based HPC systems using already available techniques demonstrated on other architectures (e.g., Intel). Three areas of related work are relevant to our contributions: sampling techniques, the impact of the ISA on application behaviour, and the growing momentum of ARM in HPC.

### A. Sampling Techniques

Simulation is costly in terms of both time and resources [3]. Hence, there is a large body of work on reducing the amount of simulation time needed to predict the performance of a workload using sampling techniques. In this section, we present an overview of common techniques.

Sherwood *et al.* presented SimPoint, a methodology that uses Basic Block Vectors (BBV) to capture program behaviour and a clustering algorithm to analyse and select representative phases for simulation [4]. Wunderlich *et al.* approach the selection of benchmark phases by systematically sampling simulations, achieving high accuracy by simulating many small portions of applications [5]. Sunwoo *et al.* show that the SimPoint methodology accurately predicts the CPI (Cycles Per Instruction) of full, single-threaded, mobile workloads [8]. In contrast to [8], our work is *a cross-architectural evaluation focusing on real hardware, as opposed to simulation*. Additionally, the techniques presented in [4], [5], [8] are well-established for single-threaded workloads, but do not consider the additional requirements of multi-threaded applications, such as thread synchronisation.

Our work focuses specifically on OpenMP-based workloads, as a large class of HPC applications falls into this category (see Table I for a list of workloads we use for evaluation). In this context, Carlson *et al.* propose a time-based sampling methodology, which allows multi-threaded workloads to be sampled with high accuracy, i.e., sub-3% error [6]. One key limitation of this technique is that it requires functional simulation of the entire program, including cache warming and detailed synchronisation between the simulation of samples. As a follow-up, using the SimPoint tool, Carlson *et al.* [7] propose the BarrierPoint methodology, which selects representative regions in OpenMP multi-threaded applications. BarrierPoint uses OpenMP barriers as natural synchronisation points to delimit applications phases. These inter-barrier regions are called barrier points. The methodology analyses the barrier points that compose each application using abstract characteristics (e.g., BBV) to determine the most representative regions. BarrierPoint offers higher speed-up than [6], because the individual barrier points can be simulated in parallel. Thus, we build upon the BarrierPoint methodology.

### B. Exploring the Impact of the ISA

Shao and Brooks propose a technique using compiler intermediate representation to perform an ISA-independent characterisation of workloads to help accelerator designers match the workloads to specialised architectures [9]. In this study we do not use ISA-independent characteristics but instead determine *whether representative regions based on abstract workload characteristics are representative when executed on a different ISA than the one used to identify them.*

Blem *et al.* conducted a study across different implementations of the ARM and x86 ISAs [10]. They conclude that the micro-architecture is the main limiting factor in terms of performance, energy, and power, and state that ARM and Intel implementations are simply different engineering design points. They also observed that the ISA effects on instruction count and mix are indistinguishable between x86 and ARM implementations. Our work follows a similar philosophy, but *we assess if a sampling methodology (BarrierPoint) can be applied more broadly than originally shown*, i.e., across x86_64 and ARMv8.

*C. ARM in HPC*

Intel has traditionally dominated the HPC market: 91% of the Top500 HPC systems are based on Intel technology [11]. With power as a main constraint, and performance-per-Watt as a design objective, ARM systems, which prevail in the embedded and mobile sectors, are getting more relevant in HPC [12].

In 2011, the 64-bit ARMv8 architecture was announced [13], meeting the demand for 64-bit addressing, among other key features required for HPC. Laurenzano *et al.* [14] show that improvements in clock rates and memory speeds achieved by 64-bit ARM systems, with respect to their 32-bit counterparts, substantially improve performance and energy-efficiency. Maqbool *et al.* found that compared to a conventional Intel x86-based cluster, the performance-to-power-ratio for HPC applications on an ARM Cortex-A9 cluster is 4 times for single core and 2.6 times for 4 cores [15]. In [16], Ou *et al.* analyse an ARM-based compute cluster and conclude that ARM is more energy-efficient for computationally lightweight applications than an Intel workstation. In summary, these works point out the potential of ARM technology for the new generation of energy-efficient HPC systems.

## III. VECTOR CAPABILITIES ON X86_64 AND ARMV8

Scientific HPC applications process immense amounts of data, which demands increased parallelisation of program execution. Vector operations (commonly referred to as Single Instruction Multiple Data, or SIMD) have gained significant traction because they exploit fine-grained data parallelism. The current trend in HPC towards more data parallelism, including larger vector lengths, increases the importance of vector extensions.

Advanced Vector Extensions (AVX) [17] are a SIMD extension on Intel processors, which introduced new instructions and increased the size of the SIMD registers from 128-bit to 256-bit. ARM introduced enhanced floating-point capabilities in the ARMv7 architecture, together with the Advanced SIMD Unit [18] (also known as ARM NEON), in its ARM Cortex-A series in 2009. ARMv8 added double-precision floating-point capabilities and increased the number of 128-bit registers from 16 to 32. These capabilities will be extended with the Scalable Vector Extension (SVE), which will allow system designers to choose the most appropriate vector length for their applications [19]. Both AVX and Advanced SIMD have an extensive set of instructions for logical and arithmetic operations, for conversion to/from vector operands, and for data movement. One of the main differences is that AVX includes 16 256-bit registers, while Advanced SIMD includes 32 128-bit registers.

TABLE I
APPLICATIONS DEPLOYED AND THEIR DESCRIPTIONS.

| | |
|---|---|
| `AMGMk` [20] | Algebraic MultiGrid Microkernel: parallel algebraic multi-grid solver for linear systems **Input:** `None` |
| `CoMD` [21] | Co-designed Molecular Dynamics: a classical molecular dynamics proxy application **Input:** `-e -T 4000` |
| `graph500` [22] | Graph500 benchmark: generation of, and Breadth first search through, an undirected graph **Input:** `-s 16` |
| `HPCG` [23] | High Performance Conjugate Gradients: preconditioned Conjugate Gradient method **Input:** `40 40 40 60` |
| `HPGMG-FV` [24] | High Performance Geometric Multigrid: a proxy application for finite volume based geometric linear solvers **Input:** `4 4` |
| `LULESH` [25] | Livermore Unstructured Lagrangian Explicit Shock Hydrodynamics **Input:** `-s 40 -i 20` |
| `MCB` [26] | Monte Carlo Benchmark: a simple heuristic transport equation using a Monte Carlo technique **Input:** `--nZonesX 200 --nZonesY 160 --numParticles 320000 --distributedSource --mirrorBoundary` |
| `miniFE` [27] | Implicit Finite Elements: a proxy application for unstructured implicit finite element codes **Input:** `nx=100 ny=100 nz=100` |
| `PathFinder` [28] | Signature-search mini-application **Input:** `-x medium10.adj_list` |
| `RSBench` [29] | Monte Carlo particle transport simulation: a proxy application with a "multipole" cross section lookup algorithm **Input:** `-s small` |
| `XSBench` [30] | Monte Carlo particle transport simulation: a proxy application with macroscopic neutron cross sections **Input:** `-s small` |

Given the increasing relevance of exploiting data-level parallelism via vector extensions in the HPC market, we validate the methodology against this key architectural feature. Specifically, we examine *whether representative regions selected on an Intel platform with AVX correctly represent the workload behaviour when executed on an ARM processor with Advanced SIMD, despite differences in the two vector extensions.*

## IV. EXPERIMENTAL SETUP

In this section, we detail the applications and hardware platforms we use in our study.

*A. Applications*

We select eleven OpenMP HPC proxy- and mini-applications, as shown in Table I, to cover a range of different computational kernels, which are representative of real workloads. We run them in 1, 2, 4, or 8 thread configurations. For each application, we identify the beginning and end of the main core loop to omit initialisation and wrap-up phases, as these are not representative of the main workload behaviour. We choose the input sets such that the problem sizes do not fit in the level one and two caches, whilst ensuring that they run in a reasonable amount of time when dynamically instrumented. The problem sizes range from $5\,\text{MiB}$ up to $385\,\text{MiB}$.

TABLE II
MICRO-ARCHITECTURAL PARAMETERS OF THE INTEL AND ARM SYSTEMS.

| | |
|---|---|
| x86_64 | Intel Core i7-3770 @ 3.4 GHz (4 cores x 2 threads)<br>32 KB L1D + 32 KB L1I, 256 KB L2 per core<br>8 MB shared L3 |
| ARMv8 | ARMv8 AppliedMicro X-Gene @ 2.4 GHz (4 clusters x 2 cores)<br>32 KB L1D + 32 KB L1I per core, 256 KB L2 per cluster<br>8 MB shared L3 |

## B. Hardware Platform

To perform our measurements, we use two machines: an Intel Core i7-3770 and an ARMv8 AppliedMicro X-Gene. Table II shows more details on the micro-architectural parameters of both systems. The Intel machine is running a Linux 3.13 kernel and a 14.04.3 Ubuntu LTS release, whilst the APM X-Gene is running a 3.18.1 Linux kernel and a Debian Jessie release.

Our applications are compiled with GCC-4.8.4 for x86_64, whilst for ARMv8, we use GCC-5.1.0[1]. We use the following compiler flags:

- Non-vectorised versions
  ```
  -O2 -march=corei7-avx
  -O2 -march=armv8-a+fp
  ```
- Vectorised versions
  ```
  -O3 -march=corei7-avx -mavx
  -O3 -march=armv8-a+fp+simd
  ```

We collect statistics from the Performance Monitoring Unit (PMU). To access hardware performance counters, we use PAPI [31] on both the Intel and ARM machines. We report cycles, instructions, misses in the L1 data cache and data misses in the L2 cache. The applications we consider are proxy applications with a very small instruction footprint. Hence, we do not consider instruction misses for our estimations.

Due to limited hardware availability for both Intel and ARM platforms, we limit our study to using up to 8 threads. We leave larger scale studies for future work, as hardware becomes more commonly available.

## V. CROSS-ARCHITECTURAL EVALUATION METHODOLOGY

We extend the BarrierPoint methodology workflow [7] to compare its behaviour across multiple ISAs: we perform the characterisation of applications on the x86_64 architecture and analyse if it is possible to estimate the behaviour of those applications on both x86_64 and ARMv8. We measure the performance of barrier points using native performance counters, and then use these to reconstruct program performance, and validate against complete program execution. Next, we provide details of our methodology, present its differences with respect to the original BarrierPoint methodology, and discuss the limitations of our measurements on real hardware.

[1]The difference in version stems from the necessity to build the Pintool for barrier point discovery and analysis, which did not support GCC-5.1.0.

## A. Workflow

Our workflow consists of the following steps:

**1) Source code instrumentation:** we manually instrument the source code to delimit the beginning and end of the region of interest. This instrumentation needs to be manually inserted as only a developer will have the understanding of where the region of interest starts and ends. We also insert PAPI calls to access performance counters before each OpenMP parallel region (as needed by Step 3), which corresponds to the beginning of each barrier point [7]. The process of adding the instrumentation around the OpenMP parallel regions could be automated, however, we found it sufficiently low overhead (in the order of minutes), that we decided against automation. We build four versions of the instrumented code: two for x86_64 and two for ARMv8, one vectorised and one non-vectorised.

**2) Barrier point discovery and clustering:** this step is only run for the x86_64 versions of the binaries, as our objective is to extract the representative regions of the workloads on x86_64. By following the steps of the BarrierPoint methodology, we obtain the Basic Block Vectors (BBV) and LRU-stack Distance Vectors (LDV) of each workload using a custom Pintool [32]. Then, we combine the BBV and LDV into Signature Vectors (SV). SV are used as input to the SimPoint 3.2 clustering (k-means) [4] to select the representative regions or barrier points, such that they are sufficient to accurately predict the behaviour of the entire application. We follow suggestions given in the original BarrierPoint paper [7] for the k-means parameters. We run this process 10 times per application to account for different thread interleavings, obtaining in each case different SV characteristics, which can lead to the selection of different barrier points.

We call a selection of representative barrier points a *barrier point set*. Typically, we observe very small differences in terms of error across each set of selections (Section VI-B). The selected barrier points usually correspond to the same code regions, but different iterations of the applications' algorithms. At this stage, we also get a multiplier for each barrier point, which is a function of the barrier point's weight. Later, we use the multipliers to scale the counter values we get to estimate the performance of the entire workload.

**3) Barrier point statistic collection:** we collect statistics per barrier point using the PMU in native runs. Using performance counters on real hardware instead of cycle-accurate simulation offers several advantages. First, running on real hardware avoids any error introduced during modelling. Second, experiments on real hardware are several orders of magnitude faster. Finally, as the applications run from beginning to end, we do not need to consider warm-up issues. Nevertheless, the instrumentation of applications entails a certain overhead, and real hardware exposes the measurements to some natural variability, as will be discussed in Section V-C. To minimise that variability and eliminate unnecessary overheads, we pin the threads to processor cores to avoid thread migration, and repeat each experiment 20 times[2]. In this step we run all four binary versions (x86_64 non-vectorised, x86_64 vectorised, ARMv8 non-vectorised, ARMv8 vectorised) with performance counter instrumentation enabled on their respective platforms, in two configurations: in one, we enable the instrumentation only at the beginning and end of the region of interest, whilst in the other we also enable instrumentation of all parallel regions, to obtain statistics per barrier point.

**4) Program behaviour reconstruction:** we combine the barrier points and multipliers obtained from Step 2 and the statistics obtained by performance counters from Step 3, to estimate the entire program behaviour in terms of cycles, instructions, L1 and L2 cache data misses.

**5) Barrier point set validation:** we validate the representativeness of each barrier point set obtained in Step 2, computing the estimation error, with respect to the execution of the complete workload, based on the measurements obtained from performance counters in Step 3. If the validation is successful, the selected barrier points can be used in simulation.

As the whole workflow is run on real hardware with minimal instrumentation of the source code, the entire process typically completes within a day. This is short relative to the total simulation time, irrespective of whether the full or partial application was run. Running applications on real hardware is typically less complex than porting them to simulation.

### B. Limitations

In this paper we only discuss limitations, both cross-architectural and not, that we have encountered during our evaluation. We do not cover the limits of cross-architectural applicability, as this is an initial exploration.

Three of the eleven applications we explore contain only a single parallel region. `RSBench`, `XSBench` and `Pathfinder` are embarrassingly parallel: the core loop of each is a large parallel section and, therefore, their analysis identifies a single barrier point. By definition, that barrier point is representative on both architectures, but the methodology does not offer any potential gain in terms of simulation time, as the complete core loop would be executed. This limitation is also highlighted in the original paper [7]. For this situation, identifying ways of reducing the size of the barrier points could help, but such analysis falls out of the scope of this paper, and we leave it as future work.

Another limitation is the variability in the number of parallel regions executed. We encounter this issue with codes that iteratively converge, such as `HPGMG-FV` and `LULESH`, which can take a variable number of iterations to do so. This can change based on the number of threads (`LULESH` and `HPGMG-FV`), or when changing architectures (`HPGMG-FV`). In the former case, the variability is not an issue, assuming that the same number of parallel regions are obtained on both architectures, for the same thread count. E.g., `LULESH` executes 9,800 barrier points with 1 thread, and 9,840 when running more than 1. However, if the iteration count is architecture-dependent, the parallel sections will not match between the two architectures. As such, we cannot measure the barrier point representativeness. In the case of `HPGMG-FV`, we observe that the different number of parallel sections is due to floating-point operations converging at different rates on Intel and ARM.

### C. Statistic Collection Overhead and Variability

Our measurements using performance counters are potentially affected by two aspects: variability, due to running the applications on real hardware, and overhead, due to the instrumentation to access the performance counter registers. In the case of high variability or significant overheads, an error is introduced in the estimations.

In order to quantify the variability of our measurements, we compute the coefficient of variation of each metric for each workload configuration. `AMGMk`, `graph500`, `HPCG`, `LULESH`, `MCB` and `miniFE` show, on average, a variation of less than 1% on both Intel and ARM platforms, whilst `HPGMG-FV` presents a variation of less than 2%, except for L2 cache data miss measurements on the Intel platform, which show a variation between 3 and 9.8%. Although the variation in cycles, instructions, and L2 cache data misses for `CoMD` is also below 1%, there is a high variation in the L1 data cache miss measurements, when running on ARMv8. This variation can be as high as 57%. This is due to the low number of misses, which, when perturbed, results in a large amount of variation. In this case, running more experiments does not decrease the variation significantly. As a result, we cannot accurately estimate the L1 data cache misses when running `CoMD` on the ARM platform.

We obtain the instrumentation overhead by calculating the error of each metric when collected per barrier point, with respect to the statistics obtained without instrumentation (as explained in the Step 3 of our workflow). In general, the instrumentation adds a very small overhead to the selected metrics (between 0.1 and 2%) on both platforms. However, due to the large number and short length of barrier points in `HPGMG-FV` and `LULESH` (we go into more detail in Section VI-B), the metrics exhibit a significant error, which increases with the number of threads. Regarding the statistic collection overhead, `LULESH` shows an average overhead of 3.1%, but for some metrics this overhead increases up to 12.2%. In the case of `HPGMG-FV`, we observe an average overhead of 7.3%, but for L1 and L2 data cache misses the average overhead rises to 19.1%, and in some cases it goes over 50%. Therefore, the estimations for `LULESH` and `HPGMG-FV` are highly affected by the instrumentation overhead. To tackle this issue, we are considering methods of artificially increasing the size of barrier points above a certain threshold. However, this falls outside of the scope of this paper and we leave this as future work.

In light of the limitations of `HPGMG-FV`, we do not proceed to evaluate this application further. Without changes to either methodology or application, we do not believe our solution can be applied to `HPGMG-FV`.

---

[2]We increased the number of repetitions for a subset of our experiments observing little variation in the standard deviation; in our results, we always report the arithmetic mean and standard deviation.

## VI. EVALUATION

In this section, we present our cross-architectural analysis and our main findings. We use the PMU data from each selected barrier point to predict the total workload performance from their weighted sum, and compare that to the results obtained from measuring the entire region of interest directly. We compare the following predictions:

- **x86_64**: x86_64 non-vectorised to x86_64 non-vectorised
- **ARMv8**: x86_64 non-vectorised to ARMv8 non-vectorised
- **x86_64-vect**: x86_64 vectorised to x86_64 vectorised
- **ARMv8-vect**: x86_64 vectorised to ARMv8 vectorised

Out of the seven OpenMP applications that passed the first stages of the workflow (Section V-B), we were able to obtain estimations within a reasonable error (less than 5%) for six of them: AMGMk, CoMD, graph500, HPCG, MCB and miniFE. Although we can apply the methodology to LULESH, we do not achieve the desired results in terms of accuracy.

The rest of this section is structured as follows: we first discuss details of the methodology when applied to each application. Then, we explore the differences across the obtained barrier point sets. Finally, we analyse the accuracy of the estimations based on barrier points.

### A. Barrier Point Characteristics Discussion

We observe a range of behaviours across the applications we evaluate, and present the total number of barrier points per application, as well as the minimum and maximum number of barrier points selected by the methodology in Table III. LULESH has a very large number of barrier points (9,840) and the methodology selects between 8 and 20 to represent the workload[3]. Each barrier point corresponds to a very small region of the code: many of the barrier points correspond to the execution of less than 100,000 instructions, and the number of L2 cache data misses is in the order of 10 Misses-Per-Kilo-Instruction (MPKI). This makes it very difficult to obtain accurate estimations: too small regions are not representative of the entire application, and they are more affected by variations that appear when running on native hardware (more details in Section VI-C).

The other six applications show diverse behaviour in terms of the number of barrier points identified. We show the total number of barrier points, as well as the range of those selected by the methodology in Table III. This spread in the number of barrier points selected for each workload comes from variability when executing on real hardware. For the remainder of the evaluation, we refer to results presented in Table IV, which contains results for the barrier point sets with the lowest estimation errors for the 8-thread configurations.

As discussed in Section V-B, we encounter an issue that was already pointed out by the original paper [7]: embarrassingly parallel applications. Additionally, we highlight a new issue: very small parallel regions are problematic for accurate estimation of the full-program behaviour. These two scenarios

[3]Increasing the number of barrier points used to reconstruct the entire application behaviour did not show a significant improvement in the estimations.

TABLE III
TOTAL NUMBER OF BARRIER POINTS, AS WELL AS THE MINIMUM AND MAXIMUM NUMBER SELECTED, PER APPLICATION, ACROSS ALL CONFIGURATIONS AND BARRIER POINT DISCOVERY RUNS.

| Application | Total | Min | Max |
|---|---|---|---|
| AMGMk | 1000 | 3 | 12 |
| CoMD | 810 | 7 | 18 |
| graph500 | 197 | 8 | 20 |
| HPCG | 803 | 12 | 19 |
| LULESH | 9840 | 8 | 20 |
| MCB | 10 | 3 | 4 |
| miniFE | 1208 | 3 | 19 |

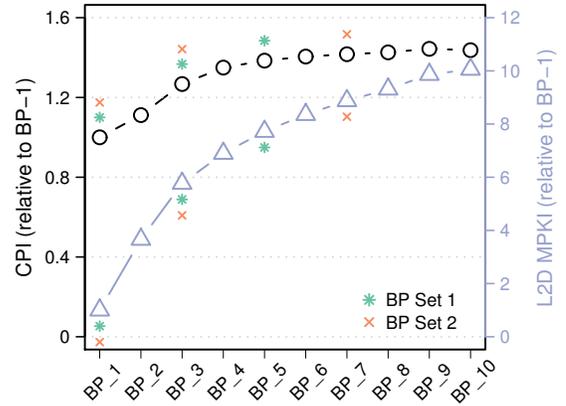

Fig. 1. Relative CPI and MPKI (with respect to BP_1) across the execution of MCB on the x86_64 platform (1-thread, non-vectorised configuration).

need to be addressed, as we found that they are relatively common in the applications we explored.

### B. Barrier Point Discovery and Selection

As explained in Section V-A, we compute 10 barrier point sets per configuration to take into account different possible thread interleavings. In general, results are consistent across the different barrier point sets, due to the regular behaviour of many HPC applications. The selected barrier points across sets usually correspond to the execution of the same parallel regions, but on different iterations of the main core loop.

Selecting a given set has a direct impact on the simulation time and resources. At the same time, sets composed out of more barrier points can lead to more accurate estimations. For example, for the miniFE vectorised configuration with 1 thread, one set contains 8 barrier points, which estimate the total number of instructions executed with less than 1% error. A different set, on the other hand, only has 3 barrier points, reducing the simulation requirements, but the estimation error for instructions executed increases to 8%. In general, although the different sets show little variation, there is a trade-off between simulation requirements (i.e., resources and time), and the estimations error we can tolerate, when selecting the more convenient barrier point set.

Not all HPC applications show such regular behaviour. A clear example is `MCB`. Figure 1 shows the normalised CPI and L2 data cache MPKI of the different regions (barrier points), with respect to the first barrier point (BP_1), for the 1-thread non-vectorised configuration. Figure 1 illustrates how the behaviour of the workload varies during different phases of its execution: the L2D MPKI significantly increases with each execution of the parallel region. This behaviour is representative of some HPC applications in which data access patterns become more irregular as the execution progresses. Both sets depicted in Figure 1 have the same number of barrier points, but result in different levels of error when estimating complete workload performance. While Set 1 is able to estimate the number of L2 misses with an error below 1%, Set 2 results in an error of 8% when estimating the same metric. We observe similar error trends on both ARMv8 and x86_64.

To summarise, selecting the most appropriate barrier point set has a significant impact on the estimation error, as well as the simulation requirements. Hence, the exploration of more barrier point sets offers a better understanding of a program's behaviour.

*C. Estimation Accuracy*

We now turn to compare the resulting accuracy for estimating x86_64 and ARMv8 full benchmark characteristics from the selected barrier points. As stated previously, we average across 20 runs for each configuration to minimise variance.

Figure 2 shows the barrier point estimation error with respect to the complete native execution for `AMGMk`, `graph500`, `HPCG`, `MCB`, `miniFE`, `CoMD` and `LULESH`. For each thread count configuration, we report the average absolute error among threads, and the maximum standard deviation. Figure 2 depicts the results for the barrier point set with the lowest error across our four metrics of interest. For each application and thread count configuration, we present the estimation error of cycles, instructions, L1 data cache misses, and L2 cache data misses, for the non-vectorised and vectorised versions of the binary, running on both Intel and ARM platforms. Table IV shows the error for the selected barrier point sets in cycles and instructions, for the 8-thread configurations. Table IV also lists the number of barrier points that compose each set, the percentage of instruction included in the representative regions, and the percentage of instructions comprising the largest barrier point in the set. This last metric gives a high-level indication of the simulation time speed-up that can be achieved, if all barrier points are executed in parallel.

In general, results are consistent on both architectures: *for the explored workloads, the abstract characteristics, represented by basic blocks and reuse distances, help us identify and select the program's representative regions for both the x86_64 and ARMv8 architectures*. Six applications (`AMGMk`, `CoMD`, `graph500`, `HPCG`, `MCB` and `miniFE`) show an error below 5% for all the metrics and configurations considered, with the exception of the L1 data misses on `CoMD`, when running on the ARM machine, and `AMGMk` L2 cache data misses for the 1-thread non-vectorised configuration, when running on both Intel and ARM platforms.

These results also extend to the vectorised versions of the workloads. With respect to the non-vectorised estimations, we do not observe any increase in the estimation error of the vectorised versions. *Despite differences in the vector extensions, the selected barrier points on Intel show similar errors on the ARM platform.*

With respect to the thread count, the error does not significantly increase when using more threads. Whilst we see an increase in the error for `graph500` and `HPCG`, the errors remain sufficiently low for the thread counts we explore. For example, for the vectorised version of `HPCG` on ARMv8, the error on L2 cache data misses increases from 0.3% with 1 thread, to 2.2% with 8 threads, but overall, the error is still very low. We also observe a larger variability when increasing the number of threads in the case of `miniFE`. However, although we have not seen any pattern of error increase with the number of threads, we observe that the standard deviation in instruction error increases for some applications (`Graph500`, `AMGMk`, `HPCG`, `miniFE` and `CoMD`). In [7], the authors show that the workload behaviour of large thread counts can be estimated using the representative regions obtained with a lower number of threads. We select the representative barrier points for each configuration independently, as our objective is to compare if those regions are representative across the two considered architectures.

The original BarrierPoint methodology includes an additional step where it computes the significance of each barrier point with respect to the total execution—based on the percentage of instructions it represents—and drops barrier points that do not significantly contribute to the total execution. We observed that following this approach to reduce the barrier point set size affects the cache estimations significantly. Given our observations and the fact that barrier points are executed in parallel, we keep all of the barrier points to obtain more accurate estimations.

`AMGMk` (Figure 2a), `HPCG` (Figure 2c) and `miniFE` (Figure 2e) exhibit a similar trend in terms of error and percentage of selected instructions (Table IV) with respect to each other—cycles and instructions can be estimated within a 2.5% error. The exception to this trend is the L2 cache data miss estimation on the 1-thread non-vectorised configuration for `AMGMk`, where the estimation errors are 8.9% and 11.0% for x86_64 and ARMv8, respectively. For the 8-thread configurations, `AMGMk` is able to accurately estimate cycles and instructions (within 2% error), while executing less than 4% of the total instructions.

Regarding `HPCG`, we observe that the estimation errors are slightly larger for ARMv8; e.g., the error in cycle estimation for the 8-thread non-vectorised configuration increases from 0.5% (x86_64) to 1.2% (ARMv8). We are able to obtain highly accurate estimations for `HPCG`, whilst executing as little as 2.7% of the workload's instructions.

`miniFE` obtains a very low error (less than 1.2%) with the selection of less than 1% of the code. Across the selected barrier points, the largest one represents a parallel region that

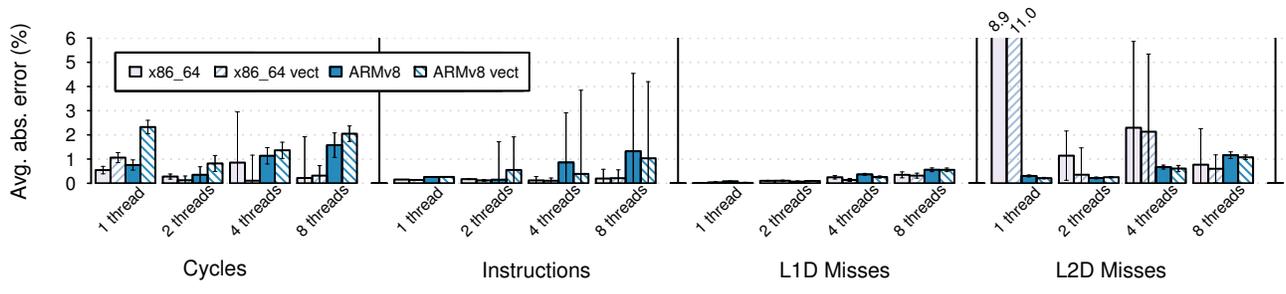
(a) AMGMk

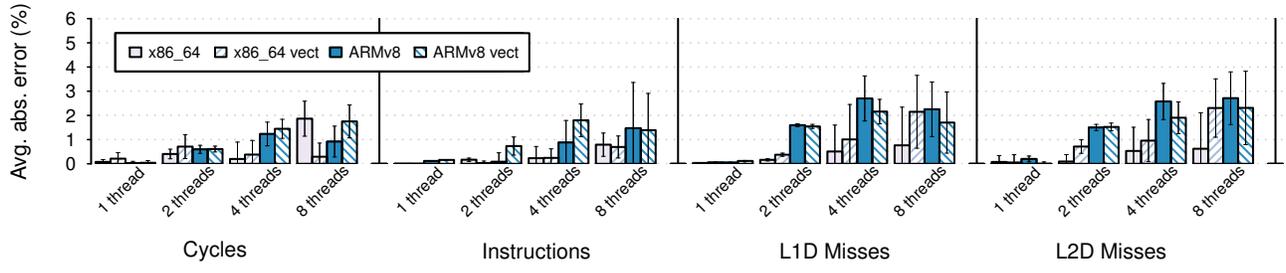
(b) Graph500

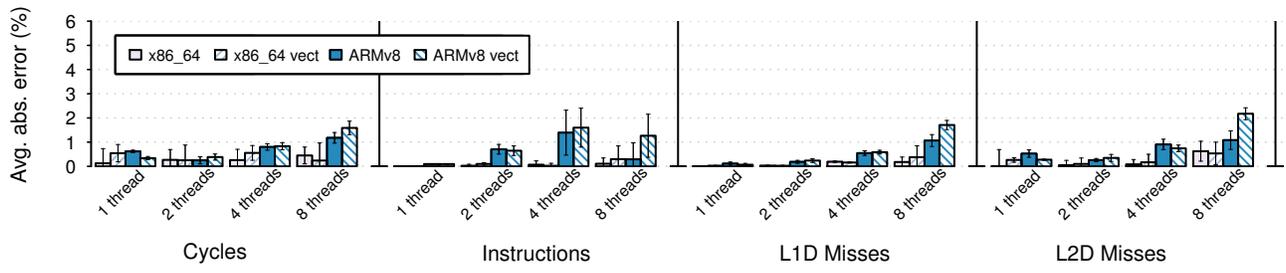
(c) HPCG

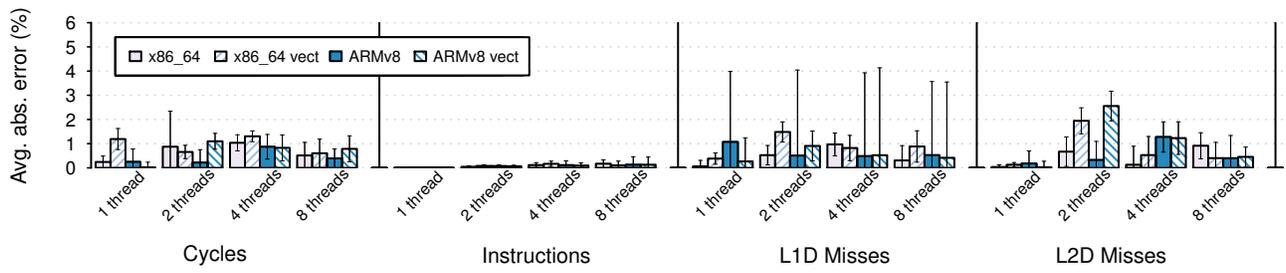
(d) MCB

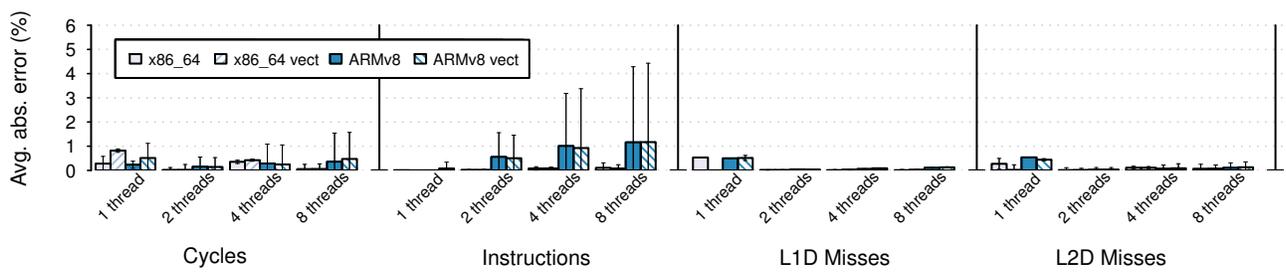
(e) miniFE

Fig. 2. Average absolute estimation error based on the x86_64 barrier points, with and without vectorisation enabled, for x86_64 and ARMv8.

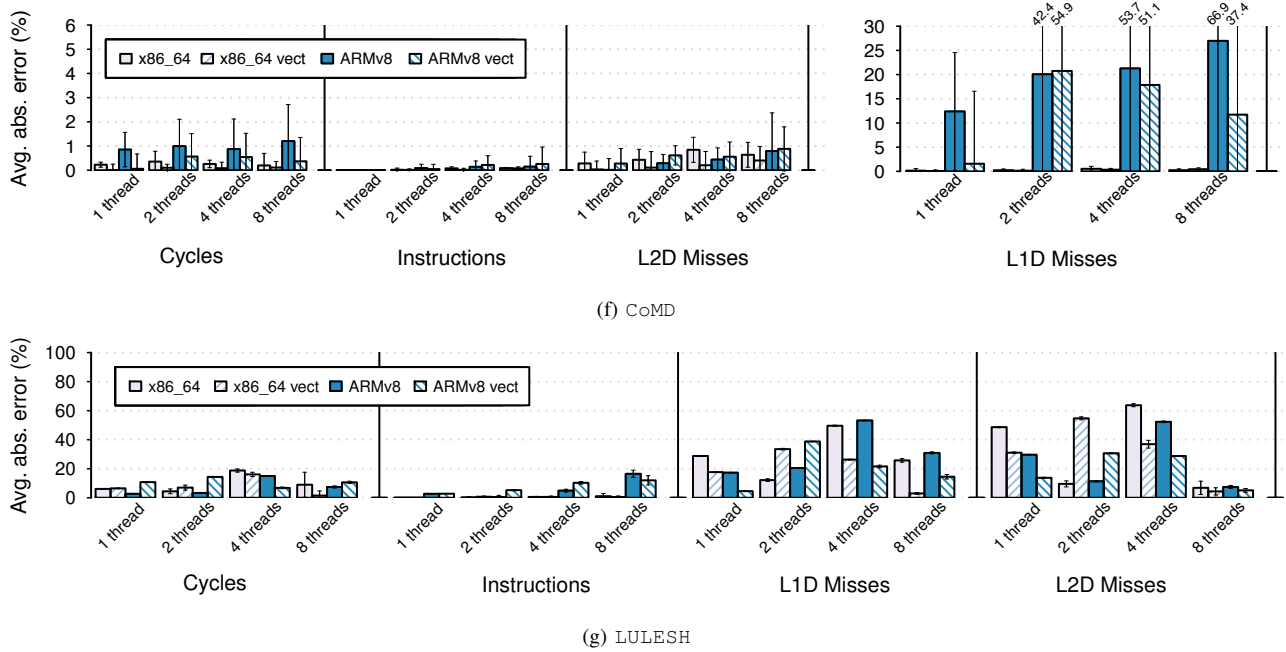

Fig. 2. Average absolute estimation error based on the x86_64 barrier points, with and without vectorisation enabled, for x86_64 and ARMv8.

dominates the execution time (85% of the instructions).

graph500 (Figure 2b) and MCB (Figure 2d) each select a large proportion of the workload. graph500 has two microkernels: graph construction and breadth first search. The graph construction is executed at the beginning of the workload and corresponds to approximately 40% of the instructions of the workload. Specifically, the function generate_kronecker_range, which generates a range of the edges of the complete graph, is run just once, but it executes around 30% of the total instructions. This function is always selected as a representative region of the workload, and therefore the maximum simulation speed-up is limited by its length (Table IV). As discussed in Section VI-B, the behaviour of MCB is irregular across the iterations of its main loop. In all cases, 3 to 4 (out of 10) barrier points are selected, forcing the significant regions to make up to 38.8% of the application's total number of instructions (Table IV).

CoMD (Figure 2f) shows a high error and variability in the L1 data cache miss estimations on the ARM platform. This illustrates the variability issues discussed on Section V-C: the high variation in the L1 data measurements makes it difficult to perform an accurate estimation of the metric, and we cannot confirm the representativeness of the barrier point set on ARMv8. For the 8-thread configuration, we report an error of less than 1.2% for cycles and instructions, while executing less than 2.1% of the application's code.

Figure 2g shows the results for LULESH. As discussed in Section V-C, the instrumentation to obtain the performance counter values introduces a significant overhead when measuring cache statistics. The error in cycles and instructions is relatively low (less than 3.5%, except for 8-thread configurations). However, the error due to the performance counter overhead does not explain the high estimation error rates by itself. For example, the overhead introduced by the instrumentation in the 8-thread configuration for measuring the cycles is less than 1%, but the estimation error reaches 17.2% for x86_64. This error also comes from the small size of the barrier points, which make difficult to estimate the entire program behaviour, and are more susceptible to perturbations while collecting statistics on real hardware.

In conclusion, for the explored applications, results are consistent across both architectures, for the non-vectorised and vectorised versions of the workloads and, in general, the error does not significantly increase when increasing the thread count. Therefore, for these applications, it is safe to run design space explorations across x86_64 and ARMv8, as we have shown that the results will be equally representative. We have also dived into several issues with the methodology. Applications with just one parallel region (barrier point) do not benefit from the methodology, and applications with too many small regions are more affected by the variability in the measurements and the instrumentation overhead, which eventually affects the estimation error. Finally, for applications whose execution depends on convergence rates, if the number of barrier points differs across architectures, it is not possible to directly estimate the application's behaviour on ARMv8 based on the x86_64 analysis.

## VII. CONCLUSIONS

We have independently demonstrated that the Barrier-Point methodology can be successfully applied to a set of HPC proxy applications across two different architectures, x86_64 and ARMv8, and across variations in architectural features, such as vector capabilities, by using AVX and the

TABLE IV
BARRIER POINTS SELECTED, ESTIMATION ERROR, AND SIMULATION SPEED-UP POTENTIAL FOR THE 8-THREAD CONFIGURATION.

| Workload | Configuration | BPs Selected[a] | Error (%) | | Instructions Selected (%) | | |
|---|---|---|---|---|---|---|---|
| | | | Cycles | Instructions | Largest BP[c] | Total[b] | Speedup[d] |
| AMGMk | x86_64 / ARMv8 | 5 / 1000 (0.5%) | 0.22 / 1.58 | 0.19 / 1.32 | 3.17 | 3.82 | 26.17x |
| | x86_64-vect / ARMv8-vect | 6 / 1000 (0.6%) | 0.32 / 2.05 | 0.21 / 1.03 | 1.79 | 2.52 | 39.68x |
| CoMD | x86_64 / ARMv8 | 17 / 810 (2.09%) | 0.20 / 1.20 | 0.09 / 0.15 | 0.52 | 2.07 | 48.30x |
| | x86_64-vect / ARMv8-vect | 12 / 810 (1.48%) | 0.11 / 0.37 | 0.08 / 0.26 | 0.55 | 1.42 | 70.42x |
| graph500 | x86_64 / ARMv8 | 10 / 197 (5.07%) | 1.86 / 0.92 | 0.79 / 1.47 | 29.27 | 38.98 | 2.56x |
| | x86_64-vect / ARMv8-vect | 9 / 197 (4.56%) | 0.29 / 1.75 | 0.70 / 1.39 | 28.55 | 38.26 | 2.61x |
| HPCG | x86_64 / ARMv8 | 17 / 803 (2.11%) | 0.45 / 1.18 | 0.11 / 0.29 | 0.63 | 2.76 | 36.23x |
| | x86_64-vect / ARMv8-vect | 12 / 803 (1.49%) | 0.24 / 1.59 | 0.30 / 1.26 | 0.62 | 1.14 | 87.71x |
| LULESH | x86_64 / ARMv8 | 10 / 9840 (0.10%) | 8.97 / 7.42 | 1.06 / 16.49 | 1.07 | 1.70 | 58.82x |
| | x86_64-vect / ARMv8-vect | 20 / 9840 (0.20%) | 1.52 / 10.60 | 0.40 / 11.99 | 0.83 | 2.37 | 42.19x |
| MCB | x86_64 / ARMv8 | 4 / 10 (40%) | 0.51 / 0.39 | 0.17 / 0.13 | 10.40 | 38.80 | 2.57x |
| | x86_64-vect / ARMv8-vect | 3 / 10 (30%) | 0.60 / 0.79 | 0.10 / 0.13 | 10.40 | 28.68 | 3.48x |
| miniFE | x86_64 / ARMv8 | 9 / 1208 (0.74%) | 0.05 / 0.36 | 0.11 / 1.16 | 0.43 | 0.56 | 178.57x |
| | x86_64-vect / ARMv8-vect | 13 / 1208 (1.07%) | 0.06 / 0.47 | 0.08 / 1.17 | 0.45 | 0.59 | 169.49x |

[a] Number of barrier points selected / Total number of barrier points (Percentage of barrier points selected). Please note that the percentage of barrier points selected does not directly correlate with the percentage of instructions executed, as the barrier points vary in size.
[b] Percentage of instructions selected (barrier point set) with respect to the workload's total instructions executed.
[c] Percentage of instructions of the largest barrier point with respect to the workload's total instructions executed (maximum simulation speed-up).
[d] Speedup as a function of reduction in the number of total instructions selected.

Advanced SIMD Unit. This enables the native profiling of applications on one architecture and the safe gathering of results on different architectures. All of our experiments, unlike prior work, which was done in simulation, have been run on native hardware. Our results show that we can identify representative sections on x86_64 and validate them on both x86_64 and ARMv8, whilst obtaining overall errors for cycles and instructions of under 2% on Intel and 2.5% on ARM, across both vectorised and non-vectorised versions. L1 and L2 data cache miss errors are also within 3%, with the exception of CoMD, which has a high variability on ARM and shows errors of up to about 25% for L1 misses estimations, and AMGMk, for the 1-thread configurations on both x86_64 and ARMv8, where error increases to 11% when estimating L2 cache data misses. In the cases where there are short barrier points, or low event counts, we recognise that small perturbations can cause large estimation errors.

We have also shown that, for the evaluated applications, it was possible to reduce the executed instructions to as little as 0.6% of the total application to obtain representative results, with a maximum of 39%. In the context of simulation, such a significant reduction enables designers and researchers to run a wider range of applications, and to use larger, more representative working sets as inputs.

The BarrierPoint methodology has several challenges. First, it cannot be applied to every OpenMP application. Applications with too few parallel regions (XSBench, RSBench and Pathfinder) do not achieve the desired simulation speed-up, and applications with too many short parallel regions (HPGMG-FV and LULESH) show high estimation error due to variability. Second, we observed that due to the natural variability of multi-threaded applications on native hardware, where different thread interleavings are possible, we obtain different abstract characteristics that lead to several barrier point sets. We showed that these sets present variability in terms of estimation error and simulation speed-up.

## VIII. FUTURE WORK

In the future, we plan on looking at extensions to this work, such that we can broaden its applicability to more workloads. A few of the opportunities we would like to explore are:
- Evaluating the applicability of the methodology across different core types, such as in-order versus out-of-order.
- Validating the representative sections against a more comprehensive set of performance counters.
- Adjusting the size of barrier points so that more applications benefit from the BarrierPoint methodology, such as RSBench, XSBench, and LULESH.
- Generalising the methodology to work on non-OpenMP applications.
- Quantifying cross-architectural ISA differences, and explore the methodology's cross-architectural applicability limits.

## ACKNOWLEDGEMENTS

The authors would like to thank Stephan Diestelhorst, Chris Adeniyi-Jones, Eric Van Hensbergen, Jonathan Beard and Charles García-Tobin for their feedback and support at the different stages of this paper, and the anonymous reviewers for their valuable feedback.

Ferrerón was supported in part by grants gaZ: T48 research group (Aragón Gov. and European ESF), TIN2013-46957-C2-1-P, TIN2016-76635-C2-1-R, Consolider NoE TIN2014-

52608-REDC (Spanish Gov.) and HiPEAC-3 NoE (European FET FP7/ICT 287759). Ruşitoru has received funding from the European Union's Horizon 2020 research and innovation programme under grant agreement N° 671697.## REFERENCES

[1] DOE ASCAC Subcommittee, "Top ten exascale research challenges," U.S. Department of Energy Office of Science, Tech. Rep., 2014. Available: http://science.energy.gov/ /media/ascr/ascac/pdf/meetings/20140210/Top10reportFEB14.pdf

[2] N. Binkert et al., "The gem5 simulator," *SIGARCH Comput. Archit. News*, vol. 39, no. 2, pp. 1–7, Aug. 2011.

[3] A. Sandberg et al., "Full speed ahead: Detailed architectural simulation at near-native speed," in *IEEE Int. Symp. on Workload Characterization*, 2015, pp. 183–192.

[4] T. Sherwood et al., "Automatically characterizing large scale program behavior," in *10th Int. Conf. on Architectural Support for Programming Languages and Operating Systems*, 2002, pp. 45–57.

[5] R. E. Wunderlich et al., "SMARTS: Accelerating microarchitecture simulation via rigorous statistical sampling," in *30th Annual Int. Symp. on Computer Architecture*, 2003, pp. 84–97.

[6] T. Carlson et al., "Sampled simulation of multi-threaded applications," in *IEEE Int. Symp. on Performance Analysis of Systems and Software*, 2013, pp. 2–12.

[7] ——, "BarrierPoint: Sampled simulation of multi-threaded applications," in *IEEE Int. Symp. on Performance Analysis of Systems and Software*, 2014, pp. 2–12.

[8] D. Sunwoo et al., "A structured approach to the simulation, analysis and characterization of smartphone applications," in *IEEE Int. Symp. on Workload Characterization*, 2013, pp. 113–122.

[9] Y. S. Shao et al., "ISA-independent workload characterization and its implications for specialized architectures," in *IEEE Int. Symp. on Performance Analysis of Systems and Software*, 2013, pp. 245–255.

[10] E. Blem et al., "Power struggles: Revisiting the RISC vs. CISC debate on contemporary ARM and x86 architectures," in *IEEE 19th Int. Symp. on High Performance Computer Architecture*, 2013, pp. 1–12.

[11] "Top500 Supercomputer Sites," http://www.top500.org/.

[12] M. Feldman, "Fujitsu switches horses for Post-K supercomputer, will ride ARM into exascale." Available: https://www.top500.org/news/fujitsu-switches-horses-for-post-k-supercomputer-will-ride-arm-into-exascale/

[13] "ARM discloses technical details of the next version of the ARM architecture," http://www.arm.com/about/newsroom/arm-discloses-technical-details-of-the-next-version-of-the-arm-architecture.php, retrieved May 2016.

[14] M. A. Laurenzano et al., "Characterizing the performance-energy tradeoff of small ARM cores in HPC computation," in *20th Euro-Par Int. Conf. on Parallel Processing*, 2014, pp. 124–137.

[15] J. Maqbool et al., "Evaluating ARM HPC clusters for scientific workloads," *Concurrency and Computation: Practice and Experience*, vol. 27, no. 17, pp. 5390–5410, Jul. 2015.

[16] Z. Ou et al., "Energy- and cost-efficiency analysis of ARM-based clusters," in *12th IEEE/ACM Int. Symp. on Cluster, Cloud and Grid Computing*, 2012, pp. 115–123.

[17] Intel Inc., "Introduction to Intel Advanced Vector Instructions," https://software.intel.com/en-us/articles/introduction-to-intel-advanced-vector-extensions.

[18] ARM Ltd., "ARM Neon SIMD Engine," http://www.arm.com/products/processors/technologies/neon.php.

[19] N. Stephens, "ARMv8-A Next Generation Vector Architecture for HPC," in *Hot Chips: A Symposium on High Performance Chips*, 2016.

[20] Lawrence Livermore National Labs. (2008) ASC sequoia benchmark codes. Available: https://asc.llnl.gov/sequoia/benchmarks/

[21] ExMatEx. (2013) CoMD: Classical molecular dynamics proxy application. Available: https://github.com/exmatex/CoMD

[22] R. C. Murphy et al., "Introducing the graph 500," *Cray Users Group (CUG)*, 2010. Available: https://github.com/graph500/graph500

[23] J. Dongarra et al., "HPCG benchmark: a new metric for ranking high performance computing systems," University of Tennessee, Knoxville, Tech. Rep. UT-EECS-15-736, 2015. Available: http://www.hpcg-benchmark.org/downloads/hpcg-3.0.tar.gz

[24] M. F. Adams et al., "HPGMG 1.0: A benchmark for ranking high performance computing systems," Lawrence Livermore National Lab, Tech. Rep. LBNL-6630E, 2014. Available: https://bitbucket.org/hpgmg/hpgmg

[25] I. Karlin et al., "Exploring traditional and emerging parallel programming models using a proxy application," in *27th IEEE Int. Parallel and Distributed Processing Symp.*, 2013, pp. 1–14. Available: https://codesign.llnl.gov/lulesh/lulesh2.0.3.tgz

[26] Lawrence Livermore National Lab. (2013) Monte Carlo Benchmark. Available: https://asc.llnl.gov/CORAL-benchmarks/Throughput/mcb-20130723.tar.gz

[27] M. A. Heroux et al., "Improving Performance via Mini-applications," Sandia National Laboratories, Tech. Rep. SAND2009-5574, 2009. Available: http://mantevo.org/downloads/miniFE_ref_2.0.html

[28] A. M. Deshpande et al., "PathFinder: A signature-search miniapp and its runtime characteristics," in *5th Workshop on Irregular Applications: Architectures and Algorithms*, 2015, pp. 9:1–9:4. Available: https://mantevo.org/downloads/PathFinder_1.0.0.html

[29] J. R. Tramm et al., "Performance analysis of a reduced data movement algorithm for neutron cross section data in Monte Carlo simulations," in *Solving Software Challenges for Exascale*. Springer, 2014, pp. 39–56. Available: https://github.com/ANL-CESAR/RSBench

[30] ——, "XSBench: the development and verification of a performance abstraction for Monte Carlo reactor analysis," *The Role of Reactor Physics toward a Sustainable Future (PHYSOR)*, 2014. Available: https://github.com/ANL-CESAR/XSBench

[31] S. Browne et al., "A scalable cross-platform infrastructure for application performance tuning using hardware counters," in *ACM/IEEE Conf. on Supercomputing*, 2000, pp. 42–42.

[32] C.-K. Luk et al., "Pin: Building customized program analysis tools with dynamic instrumentation," in *ACM SIGPLAN Conf. on Programming Language Design and Implementation*, 2005, pp. 190–200.